# Degradation of Phosphorene in Air: Understanding at Atomic Level

Gaoxue Wang[1], William J. Slough[1], Ravindra Pandey[1]*, Shashi P. Karna[2]

[1]Department of Physics, Michigan Technological University, Houghton, Michigan 49931, USA
[2]US Army Research Laboratory, Weapons and Materials Research Directorate, ATTN: RDRL-WM, Aberdeen Proving Ground, MD 21005-5069, USA

(September 2, 2015)

Email: pandey@mtu.edu

**Abstract**

Phosphorene is a promising two dimensional (2D) material with a direct band gap, high carrier mobility, and anisotropic electronic properties. Phosphorene-based electronic devices, however, are found to degrade upon exposure to air. In this paper, we provide an atomic level understanding of stability of phosphorene in terms of its interaction with $O_2$ and $H_2O$. The results based on density functional theory together with first principles molecular dynamics calculations show that $O_2$ could spontaneously dissociate on phosphorene at room temperature. $H_2O$ will not strongly interact with pristine phosphorene, however, an exothermic reaction could occur if phosphorene is first oxidized. The pathway of oxidation first followed by exothermic reaction with water is the most likely route for the chemical degradation of the phosphorene-based devices in air.

# 1.0 Introduction

Phosphorene is one of the group V elemental monolayers[1-3] with a direct band gap, high carrier mobility, and anisotropic electronic properties making it a promising candidate for applications in electronics and optoelectronics[4, 5]. The chemical degradation of phosphorene upon exposure to ambient conditions, however, is a challenge to the stability and performance of phosphorene-based devices[6-10]. The presence of oxygen and humidity is suggested to be the main cause of the degradation process[11-13]. Recent experiments have also demonstrated the photo-assisted degradation of phosphorene[14], which is predicted to be related to intrinsic defects[15]. Despite experimental and theoretical efforts, there are still some unanswered questions regarding the degradation of phosphorene, including (i) atomic level of understanding on the degradation process of phosphorene; (ii) the role of $H_2O$ in the degradation process; and (3) the environmental stability of other phosphorene allotropes (e.g., blue phosphorene[16]).

In order to address these questions, we have performed density functional theory (DFT) calculations combined with first principles molecular dynamics (MD) simulations to investigate the interactions of $O_2$ and $H_2O$ with phosphorene. In Sec. 3.1, we focus on the interaction of $O_2$ with phosphorene. Since surface reaction with $O_2$ has been reported to be crucial in the degradation process of black phosphorene[13], we will extend the discussion to blue phosphorene. In Sec. 3.2, the adsorption of $H_2O$ on phosphorene allotropes is investigated in terms of adsorption configuration, binding energy, and bonding characteristics. In Sec. 3.3, we discuss the degradation of phosphorene by calculating the relative energies along a likely interaction pathway. Our calculated results show that $O_2$ can spontaneously dissociate on phosphorene at room temperature; $H_2O$ will not strongly interact with pristine phosphorene, however, an exothermic reaction could occur if phosphorene is first oxidized. Other allotropes of phosphorene, e.g. blue phosphorene are also expected to deteriorate in air.

# 2.0 Computational details

The electronic structure calculations were performed using the Vienna *ab initio* simulation package (VASP)[17, 18]. The exchange-correlation was treated within the generalized gradient approximation (GGA) using Perdew−Burke−Ernzerhof (PBE)[19] functional for the calculations. We also employed the DFT-D2 method of Grimme[20] to include contributions from the van der Waals (vdW) interactions. The energy of convergence was set to $1 \times 10^{-6}$ eV and the residual force on each atom was smaller than 0.01 eV/Å during the structural optimization. The cutoff energy for the plane-wave basis was set to 500 eV. The vacuum distance normal to the plane was larger than 30 Å to eliminate interaction between the periodic replicas. A rectangular supercell of (3×4) was used for the black phosphorene, and a parallelogram supercell of (4×4) was used for the blue phosphorene. The reciprocal space was sampled by a grid of (2×2×1) $k$ points in the Brillouin Zone.

First principles molecular dynamics (MD) simulations were also performed to simulate the interaction processes considered. The MD simulation was based on the norm-conserving Troullier-Martins pseudopotential together with Nosé thermostat[21] as implemented in the SIESTA program package[22]. In order to minimize the constraints induced by periodicity in the slab model for MD simulations, (5×6) and (5×5) supercells were used for black and blue phosphorene, respectively. The time step was set to 1 $fs$, and the simulation temperature was set to 300 K.

### 3.0 Results and Discussion

Black phosphorene has a puckered surface with two sub-layers of phosphorus atoms which are arranged in a rectangular lattice. At GGA-PBE level of theory, the lattice constants along the armchair and the zigzag direction are 4.57 Å and 3.31 Å, respectively. The bond lengths are 2.22 Å and 2.25 Å. Blue phosphorene has a buckled honeycomb structure with lattice constant of 3.28 Å and bond length of 2.26 Å. Our results are in agreement with the reported lattice constants and bond lengths of black and blue phosphorene[2, 16], thereby, demonstrating the reliability of the modeling elements used in the calculations.

### 3.1 O$_2$ interacting with phosphorene

It has been established by both theoretical calculations and experiments that O$_2$ can easily dissociate on black phosphorene[13, 23] leading to the formation the oxidized lattice[24]. The exothermic energy ($\Delta Q$) of O$_2$ dissociation on black phosphorene is 4.46

eV/$O_2$ molecule, and that on blue phosphorene is 4.35 eV/$O_2$ molecule (Figure 1). The details of calculations for $\Delta Q$ are given in Table S1 in the supplementary materials.

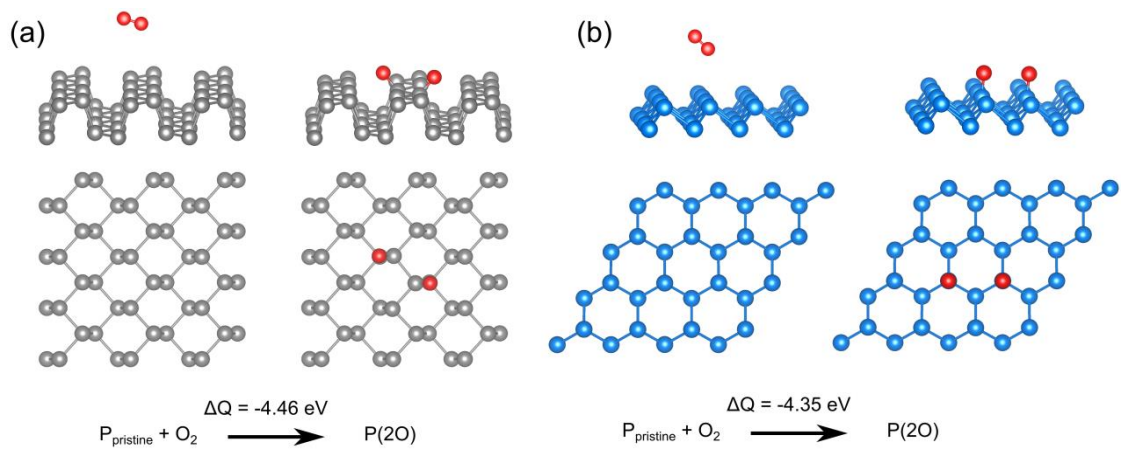

*Figure 1. $O_2$ dissociation on phosphorene: (a) black phosphorene, (b) blue phosphroene. P(2O) represents black or blue phosphorene with two O adatoms.*

After dissociation, atomic oxygen finds the dangling position to be the preferred site on black phosphorene which is consistent with previous theoretical studies[13, 24, 25]. This is not the case with blue phosphorene where the preferred site is the top site (See Figure S1 in the supplementary materials). The P-O bond shows similar bonding character as seen from the bond length and Bader charges[26] given in Table 1. Overall, the nature of interaction of oxygen with phosphorene stems from the $sp^3$ bonds which leave a lone electron pair on each phosphorous atom, and the preferred binding site follows the direction of the lone electron pair on both allotropes.

*Table 1. Structural properties of atomic O adsorbed on phosphorene.*

| Phosphorene | Black | Blue |
|---|---|---|
| Bond length $R_{P-O}$ (Å) | 1.50 Å | 1.50 Å |
| Bond angle $\angle$P-P-O (°) | 112°, 117° | 123° |
| Bader charge, Oxygen | -1.31e | -1.32e |

The calculated results based on first principles MD simulations further affirm the dissociation of $O_2$ on phosphorene. Figure 2 shows time-dependent snapshots of the configurations showing interaction of oxygen with phosphorene during MD simulations. These configurations were obtained by placing a few $O_2$ molecules 4 Å initially above the surface at a constant temperature of 300 K. For the case of black phosphorene, some $O_2$ molecules will first move closer to the native phosphorus atoms, then dissociate into atomic oxygen atoms after 600 *fs* (See video V1 in the supplementary materials). A similar $O_2$ dissociation process is seen on the blue phosphorene after 1150 *fs* (See video V2 in the supplementary materials). The calculated results therefore show that both allotropes of phosphorene will go through the spontaneous oxidation process upon exposure to $O_2$ at room temperature.

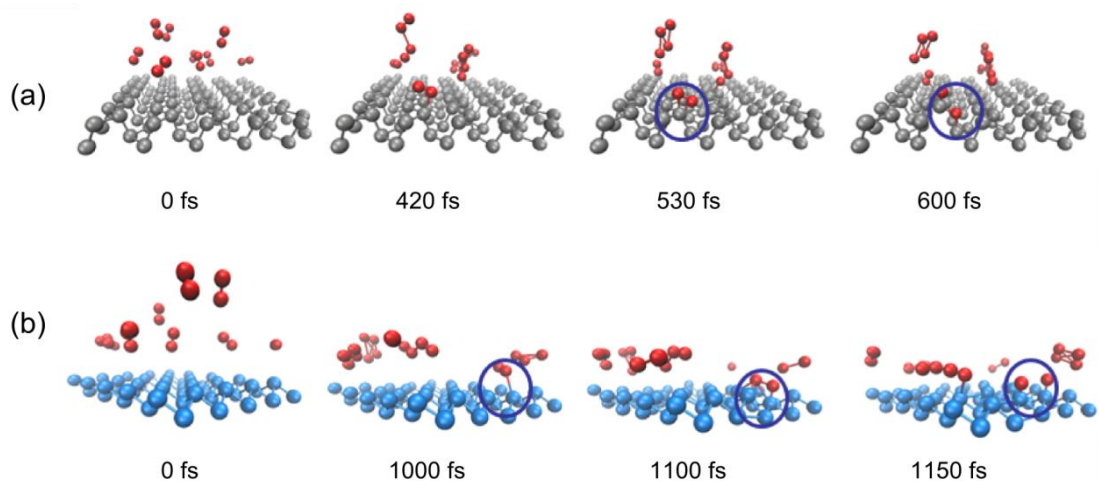

*Figure 2. Snapshots of $O_2$ interacting with phosphorene during MD simulations: (a) black phosphorene, (b) blue phosphorene.*

## 3.2 $H_2O$ interacting with phosphorene

Humidity is found to be another important factor in determining the stability of phosphorene in air. Castellanos-Gomez et al.[7] reported that $H_2O$ adsorbed on phosphorene will induce a significant distortion to the structure. Contradictory results were obtained by Baisheng et al.[27], and Yongqing et al.[28] suggesting that phosphorene is stable in the presence of $H_2O$. Since the effect of $H_2O$ on the stability of phosphorene is still ambiguous, we have now performed DFT calculations on $H_2O$ interacting with phosphorene.

Figure 3 shows the configurations of $H_2O$ interacting with phosphorene considered for the calculations: one leg, two leg, and O closer. The configuration referred to as "one leg" is the configuration in which one of the H atoms is closer to the surface, "two leg" means both H atoms are closer to the surface, and "O closer" means the O atom is closer to the surface.

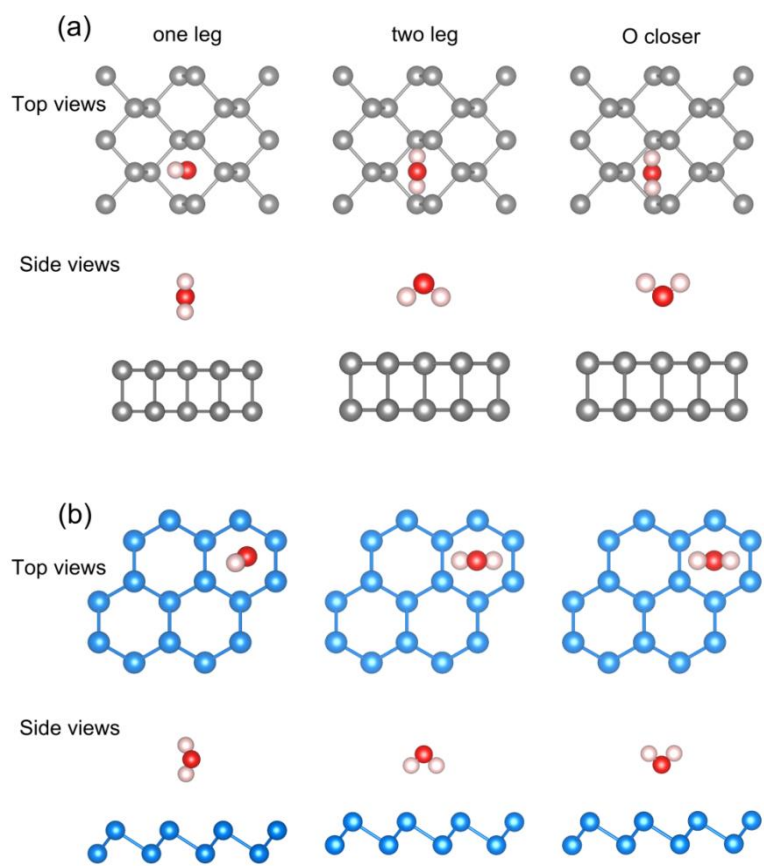

*Figure 3. Top and side views of the configurations considered for $H_2O$ interacting with phosphorene: (a) black phosphorene, (b) blue phosphorene.*

The calculated binding energy profiles with vdW correction using DFT-D2 method of Grimme[20] are shown in Figure 4. Some of the results deduced are:

(i). The 'two leg' configuration is the most stable configuration suggesting that H atoms prefer to move towards the surface. This is due to the well-known polar nature of the $H_2O$ molecule in which H atoms tend to attract the lone electron pairs of phosphorene.

(ii). The GGA-PBE functional underestimates the binding energy for all the cases (See Figure S2 in the supplementary materials). This is due to the fact that GGA-PBE functional fails in correctly accounting for the vdW interaction, which plays a significant role in the adsorption of molecules on a substrate[29].

(iii). The calculated binding energy is about 180 and 125 meV for $H_2O$ on the black and blue phosphorene, respectively with vdW correction. It is larger than that of $H_2O$ on graphene at the same level of theory (in the range from 60 to 120 meV [29]), which is mainly due to the presence of the lone electron pairs on phosphorene. The attraction between H atoms and the lone electron pairs of phosphorene is clearly demonstrated by the deformation charge density plots (See Figure S3 in the supplementary materials) showing the increase in the electron density in the region between H and P atoms.

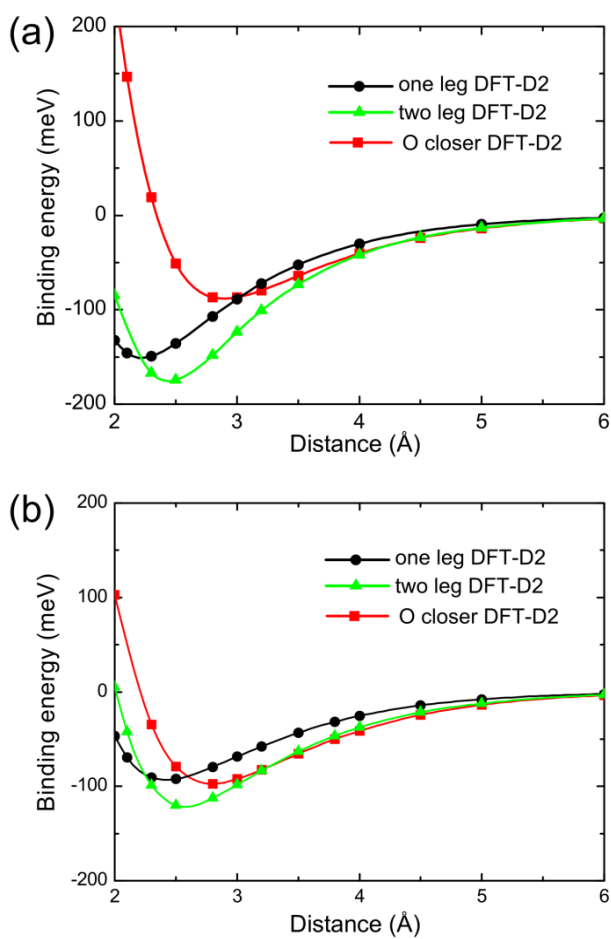

*Figure 4. The calculated binding energy profiles of a $H_2O$ molecule approaching phosphorene: (a) black phosphorene, (b) blue phosphorene.*

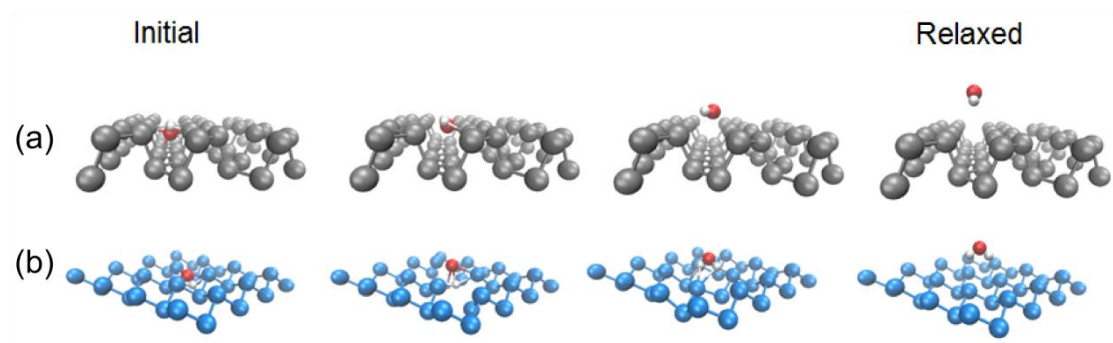

*Figure 5. Relaxation process of a $H_2O$ moleclue initially "forced"into phosphorene lattice: (a) black phosphorene, (b) blue phosphorene. The $H_2O$ moves out the lattice after structural relaxiation.*

In order to further examine the interaction of $H_2O$ with phosphorene, we considered the initial configuration to consist of a "forced" $H_2O$ molecule at the interstitial site of the phosphorene lattice. If $H_2O$ prefers to interact strongly with phosphorene, then the optimized configuration should show that H and O atoms remain in the lattice. This is not the case as $H_2O$ moves out of the lattice to a surface site (Figure 5) without distorting the surface for both allotropes. Our first principles MD simulations up to 10 *ps* also find that $H_2O$ molecules stay near the phosphorene surface without any chemical interaction within 10 *ps* (See video V3 in the supplementary materials). Therefore, instead of strongly interacting with phosphorene, $H_2O$ prefers to bind to the surface though hydrogen bonds.

**3.3 Stability of phosphorene in Air : Exposure to $O_2$ and $H_2O$**

Based on the calculated results giving in Sec. 3.1 and Sec. 3.2, we can state that the $O_2$ molecule could spontaneously dissociate on phosphorene, and the $H_2O$ molecule will not interact strongly with the pristine phosphorene.

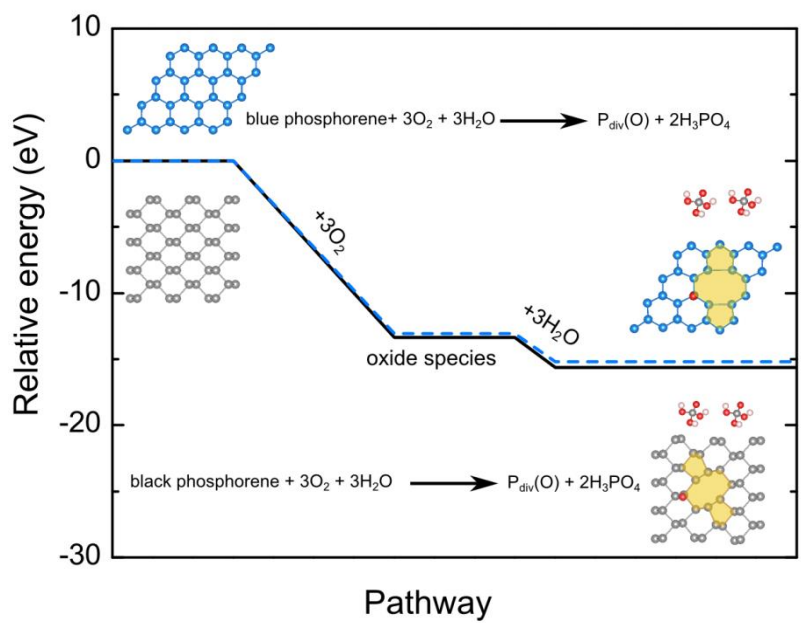

*Figure 6. Relative energy during the interaction process of black (solid curve) and blue (dashed curve) phosphorene with $O_2$ and $H_2O$. The insets show the structure of initial phosphorene structure and the products. $P_{div}(O)$ represents defective phosphorene with one divacancy and one O adatom.*

Considering that the phosphorous oxides (e.g., $P_3O_6$, $P_4O_{10}$) are reactive with $H_2O$[30], we offer a simple intuitive view of the degradation of phosphorene in air based on energetic considerations: first, oxidation of the 2D lattice of phosphorene will occur in air; then, the oxide species will interact with $H_2O$. In order to validate our view of this interaction process, the relative energy of the initial and final chemical species along the pathway are calculated. The reactants are phosphorene, $O_2$ molecules, and $H_2O$ molecules and the products are phosphoric acid and phosphorene with defects including di-vacancy and O adatom (Figure 6). For black phosphorene, the total energy release during this process is 15.6 eV. The oxidation process is exothermic with energy release of about 13.4 eV, and the reaction of phosphorene oxide species with $H_2O$ molecule further releases 2.25 eV of energy (See Table S2 and S3 in the supplementary materials for the details of the calculations). The energy profile for blue phosphorene almost overlaps with that of black phosphorene, which suggests the similarity of black and blue phases in terms of the environmental stability. Overall, the exothermic process implies that $H_2O$ will react with phosphorene if it is oxidized on the surface. The proposed pathway will lead to the formation of phosphoric acid and defective phosphorene. The defective phosphorene could further be photo-oxidized[15], and then the oxide species will further react with $H_2O$. This reaction circle results in the fast degradation of phosphorene in air.

**4.0 Summary**

In order to investigate the stability of phosphorene in air, the interaction of $O_2$ and $H_2O$ with phosphorene was studied by using density functional theory combined with first-principles molecular dynamics simulations. We find that (i) $O_2$ will spontaneously dissociate on phosphorene at room temperature. The exposure of phosphorene to $O_2$ will induce its oxidation forming an oxidized phosphorene lattice; (ii) $H_2O$ does not interact directly (chemically) with the pristine phosphorene lattice. It prefers to bind to the surface of phosphorene though hydrogen bonds; (iii) $H_2O$ will exothermically interact with phosphorene if it has first been oxidized; (iv) Other theoretically predicted 2D phosphorene allotropes, e.g. blue phosphorene, are also expected to be unstable in air.

Our calculations are supported by several experimental results; e.g. experiments have shown that fast degradation of phosphorene occurs with the existence of both $O_2$ and $H_2O$, the degradation process slows down with the exposure of phosphorene to only $O_2$ or $H_2O$[14],[31]; experiments have shown a drop of pH after water addition to phosphorene[12], which is a clear identification of the formation of phosphoric acid. Considering the rapid growth of research on 2D materials based on the group V semiconductors, our study provides an atomic scale understanding of the stability of phosphorene in air, which will aid in determining the degradation and aging effects of phosphorene-based devices.

**Acknoledgements**

Helpful discussions with Dr. V. Swaminathan are acknowledged. Financial support from ARL W911NF-14-2-0088 is obtained. RAMA and Superior, high performance computing clusters at Michigan Technological University, were used in obtaining results presented in this paper.

**Supporting Information**

**Degradation of Phosphorene in Air: Understanding at Atomic Level**

Gaoxue Wang[1], William J. Slough[1], Ravindra Pandey[1]∗, and Shashi P. Karna[2]

[1]Department of Physics, Michigan Technological University, Houghton, Michigan 49931, USA
[2]US Army Research Laboratory, Weapons and Materials Research Directorate, ATTN: RDRL-WM, Aberdeen Proving Ground, MD 21005-5069, USA

Email:   pandey@mtu.edu

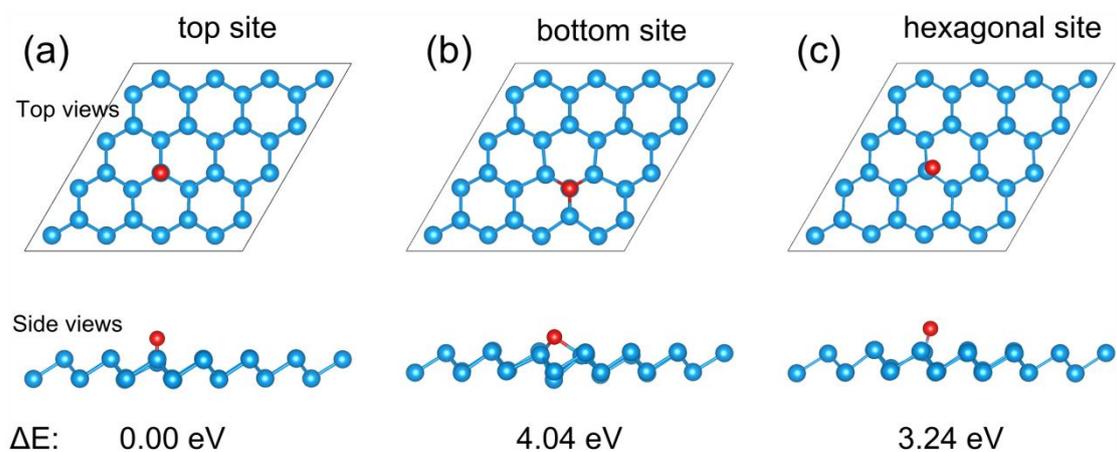

*Figure S1. Optimized structure of O atom on different sites of blue phosphorene. Note that the most stable adsorption site is the top site of phosphroene. The O atom moves away from the hexagonal site after structural optimization. ΔE is the total energy difference of different adsorption sites to that of the most stable site.*

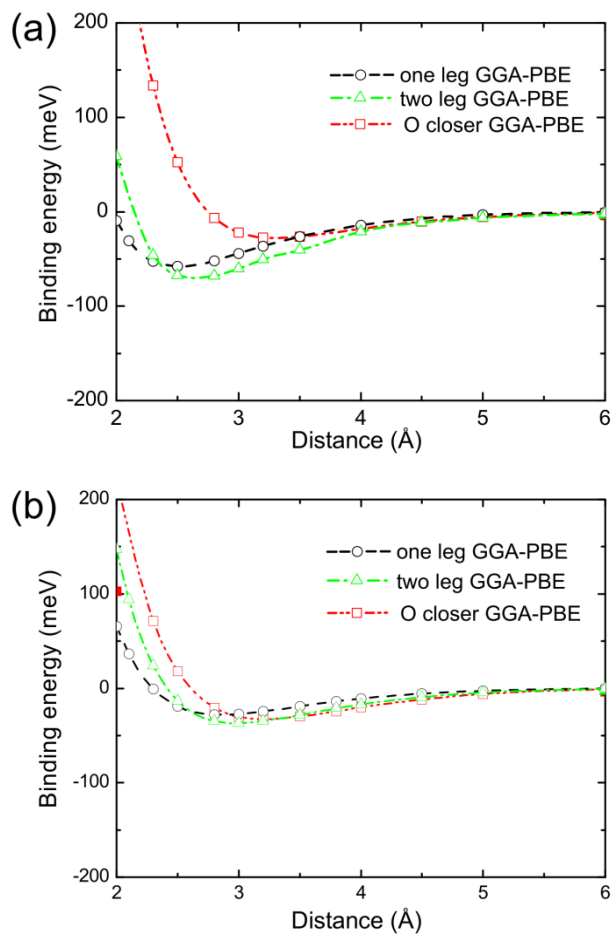

*Figure S2. The calculated binding energy profiles of a $H_2O$ molecule interacting with phosphorene at GGA-PBE functional level of theory: (a) black phosphorene, (b) blue phosphorene.*

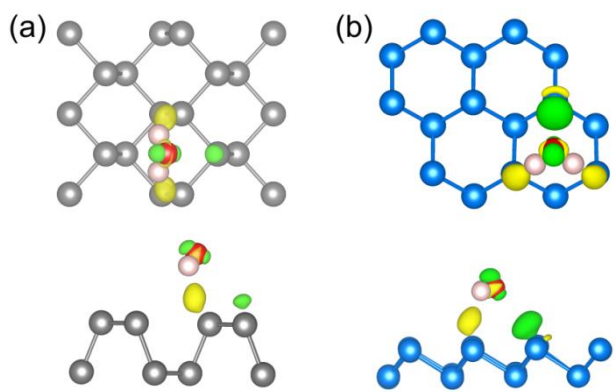

*Figure S3. Top and side views of deformation charge density of $H_2O$ molecule on: (a) black phosphorene, and (b) blue phosphorene. The yellow (green) region represents accumulation (depletion) of electrons. The isovalue is 0.001 $e/bohr^3$.*

Table S1. Energy release after $O_2$ dissociation on black and blue phosphorene. P(2O) represents black or blue phosphorene with two O adatoms. The energy is in unit of eV.

| Black phosphorene (3×4 supercell) | -262.022 | Blue phosphorene (4×4 supercell) | -171.579 |
|---|---|---|---|
| $O_2$ | -9.858 | $O_2$ | -9.858 |
| P(2O) | -276.34 | P(2O) | -185.787 |
| *ΔQ* | **-4.46** | *ΔQ* | **-4.35** |

Table S2. Energy release during the reaction of black phosphroene with $H_2O$ and $O_2$. $P_{div}(O)$ represents defective black phosphorene with a divacancy and one O adatom. The energy is in unit of eV.

| Reactants | | Products | |
|---|---|---|---|
| Black phosphorene (3×4 supercell) | -262.022 | $P_{div}(O)$ | -256.907 |
| 3 $O_2$ | -29.574 | 2 $H_3PO_4$ | -92.928 |
| 3 $H_2O$ | -42.625 | | |
| Total | -334.221 | Total | -349.835 |
| *ΔQ* = -15.614 | | | |

Table S3. Energy release during the reaction of blue phosphroene with $H_2O$ and $O_2$. $P_{div}(O)$ represents defective blue phosphorene with a divacancy and one O adatom. The energy is in unit of eV.

| Reactants | | Products | |
|---|---|---|---|
| Blue phosphorene (4×4 supercell) | -171.579 | $P_{div}(O)$ | -166.032 |
| 3 $O_2$ | -29.574 | 2 $H_3PO_4$ | -92.928 |
| 3 $H_2O$ | -42.625 | | |
| Total | -243.778 | Total | -258.96 |
| *ΔQ* = -15.182 | | | |